\documentclass[aps,twocolumn,pra,superscriptaddress,amsmath,amssymb]{revtex4-1} 
\usepackage{dcolumn}
\usepackage{bm}
\usepackage{amsmath}
\usepackage{txfonts}
\usepackage[T1]{fontenc}
\usepackage{xspace}
\usepackage{ulem}
\newcommand{\means}[1]{\langle#1\rangle}
\newcommand{\AFOAFM}{AFO($\eta$)-AFM\xspace}
\newcommand{\AFOCDW}{AFO($\xi$)-CDW\xspace}
\newcommand{\gAFO}{genuine AFO($\eta$)\xspace}

\ifx\pdfoutput\undefined
\usepackage[dvipdfmx]{graphicx}
\usepackage[dvipdfmx]{hyperref}
\usepackage[dvipdfmx]{color}
\else
\usepackage{graphicx}
\usepackage{hyperref}
\usepackage{color}
\fi

\usepackage{natbib}

\usepackage{soul}
\soulregister\cite7
\soulregister\ref7
\soulregister\pageref7

\DeclareRobustCommand{\nchange}[2]{\ifmmode{{\textrm{\setul{}{1pt}\setstcolor{blue}\st{$\displaystyle#1$}}}}\else{{\setul{}{1pt}\setstcolor{blue}\st{#1}}}\fi\ \textcolor{blue}{#2}}

\begin{document}
\let\emph\textit

\title{
  Staggered ordered phases in the three-orbital Hubbard model}
\author{Kosuke Ishigaki}
\author{Joji Nasu}
\author{Akihisa Koga}
\affiliation{
  Department of Physics, Tokyo Institute of Technology,
  Meguro, Tokyo 152-8551, Japan
}
\author{Shintaro Hoshino}
\affiliation{
  Department of Physics, Saitama University,
  Saitama 338-8570, Japan
}
\author{Philipp Werner}
\affiliation{
  Department of Physics, University of Fribourg,
  1700 Fribourg, Switzerland
}

 \date{\today}
\begin{abstract}
We study ordered phases with broken translational symmetry in the half-filled three-orbital Hubbard model
with antiferromagnetic Hund coupling by means of dynamical mean-field theory (DMFT)
and continuous-time quantum Monte Carlo simulations.
The stability regions of the antiferro-orbital (AFO), antiferro-magnetic (AFM), and
charge density wave (CDW) states are determined by measuring the corresponding order parameters.
We introduce two symmetrically distinct AFO order parameters
and show that these 
are the primary order parameters in the phase diagram.
The CDW and AFM states appear simultaneously with these two types of AFO orders
in the weak and strong coupling region, respectively. The DMFT phase diagram is
consistent with the results obtained by the Hartree approximation and
strong-coupling perturbation theory. 
In the weak coupling regime, a nontrivial exponent $\beta=3/2$ is found for the CDW order parameter,
which is related to the coupling between the CDW and AFO orders
in the Landau theory characteristic for the three-orbital model.
We also demonstrate the existence of a metallic AFO state without any charge disproportions
and magnetic orders, which appears only at finite temperatures.
\end{abstract}
\maketitle

\section{Introduction}
Materials with multiple active orbitals 
attract much interest
since they exhibit a variety of remarkable phenomena such as, e. g.,
colossal magnetoresistance in manganites~\cite{CMR}, or
exotic superconductivity in ruthenates~\cite{Maeno} and
iron pnictides~\cite{Hosono}.
In these compounds, charge, spin, and orbital degrees of freedom are
strongly coupled with each other, which leads to the emergence of novel ordered states.
A special class of multiorbital systems are the fullerene-based
solids~\cite{Zadike1500059,Takabayashi1585,10.1038/ncomms1910,Capone2,Nomura},
which show an unconventional form of superconductivity in the vicinity of the Mott insulating state.
In these compounds, triply degenerate electronic orbitals in fullerene molecules
couple with vibration modes, resulting in a strong renormalization of the local interactions \cite{Fabrizio,Nomura2015}.
The static interorbital interactions effectively become larger than
the intraorbital interactions and a sign-inverted (antiferromagnetic) Hund coupling is realized.
This negative Hund coupling is expected to play an essential role in stabilizing
the unconventional superconductivity in these compounds.
Furthermore, an unusual Jahn-Teller metal has been identified experimentally above
the superconducting critical temperature in fullerene-based solids~\cite{Zadike1500059},
which stimulates further investigations on the properties of multiorbital systems with
large interorbital interactions and antiferromagnetic Hund coupling.

In a previous effort
to clarify how the interorbital interactions stabilize low temperature states~\cite{PhysRevLett.118.177002,Ishigaki},
we have considered the three-orbital Hubbard model,
neglecting translational-symmetry-broken phases.
We have demonstrated the existence of
spontaneously orbital-selective Mott and
orbital-selective superconducting states,
which may be relevant for understanding the low temperature properties of
the fullerene-based solids.
Moreover, a two-dimensional fulleride system has recently been investigated at zero temperature
using a variational Monte Carlo method,
and the orbital symmetry breaking has been discussed in the strong coupling region~\cite{Misawa}.
As for the solution of the three-orbital Hubbard model with spontaneous translation-symmetry breaking, the instabilities of disordered states have been investigated based
on susceptibility calculations~\cite{Hoshino2016,PhysRevLett.118.177002}.
On the other hand, we still lack the complete picture of the ordered phases
at nonzero temperatures, even in the half-filled system.
Therefore, it is instructive to discuss the 
orbital-selective
staggered ordered states, as a basis for further explorations of
the symmetry-broken states in multiorbital systems such as 
A$_3$C$_{60}$.

In this paper, we investigate charge, spin, and orbital ordered states of the half-filled three-orbital Hubbard model on the bipartite lattice
as an extension of our previous studies.
We use the dynamical mean-field (DMFT) theory~\cite{DMFT1,DMFT2,DMFT3} in combination with continous-time quantum Monte Carlo (CTQMC)
simulations~\cite{CTQMC,CTQMCREV} to clarify the temperature-dependent phase diagram of this model.
We mainly examine the electron occupancies in each orbital
to investigate the appearance of staggered ordered states such as
antiferro-orbital (AFO) and antiferro-magnetic (AFM) states.
Furthermore, we demonstrate the existence of
a charge density wave (CDW) state,
which may be unexpected in a repulsively interacting system.
We find an exotic criticality of the phase transition
between the metallic and CDW phases,
where the CDW order parameter
does not exhibit a conventional mean-field-like square root behavior.
We elucidate, using the Landau theory, that
the CDW state is accompanied by the AFO order and the
critical behavior is described by two order parameters,
where the AFO order parameter
is the primary order parameter and the CDW the secondary one.
We also employ the static mean-field approximations
for the weak and strong coupling limits
to discuss the ground-state and finite-temperature properties
for the complemenrary understanding.

The paper is organized as follows.
In Sec.~\ref{sec:model},
we introduce the three-orbital Hubbard model.
In Sec.~\ref{sec:results},
we study the stability of the staggered ordered states at low temperatures,
combining DMFT with the CTQMC impurity solver.
The phase transition in the strong coupling limit is discussed
in Sec.~\ref{sec:strong}.
A summary is given in the final section.
Appendices~\ref{sec:Landau} and~\ref{sec:comment-triple-point} are devoted to
the Landau theory and
a comment on the triple point in the finite temperature phase diagram, respectively.

\section{Model}\label{sec:model}
We consider the half-filled three-orbital Hubbard model on the
infinitely-coordinated Bethe lattice, which is described by the Hamiltonian
\begin{align}\label{3orb}
  \mathcal{H}=& \mathcal{H}_t+\mathcal{H}_U,\\
   \mathcal{H}_t= & -t\sum_{\means{i,j}\alpha\sigma}c^\dagger_{i\alpha\sigma}c_{j\alpha\sigma},\\
  \mathcal{H}_U = & U\sum_{i\alpha}n_{i\alpha\uparrow}n_{i\alpha\downarrow}
      +U'\sum_{i\sigma\alpha<\beta}n_{i\alpha\sigma}n_{i\beta\bar{\sigma}}\nonumber\\
   & +(U'-J)\sum_{i\sigma\alpha<\beta}n_{i\alpha\sigma}n_{i\beta\sigma},
\end{align}
where $c_{i\alpha\sigma}\; (c_{i\alpha\sigma}^\dag)$ is an anihilation (creation)
operator for an electron
with spin $\sigma$ $(=\uparrow,\downarrow)$ and orbital index $\alpha$ $(=\!1,2,3)$
at the $i$th site and
$n_{i\alpha\sigma}=c^\dagger_{i\alpha\sigma}c_{i\alpha\sigma}$.
$t$ is the transfer integral between nearest neighbor sites,
$U(U')$ is the intra-(inter-)band Comlomb interaction and
$J$ is the Hund coupling.
We assume the relation $U=U'+2J$ and for simplicity neglect the exchange part of
the Hund coupling and pair hopping.
The Bethe lattice with connectivity $z$ is considered for the kinetic energy term, and we take $z\rightarrow \infty$ so that the half bandwidth $D=2\sqrt{z}t$ becomes constant after rescaling of the hopping parameter $t=t^*/\sqrt{z}$.
In the present calculations, we fix the Hund coupling
as $J/U=-1/4$,
which allows us to reveal the relevant physics of
the multiorbital system with antiferromagnetic Hund coupling
at a reasonable computational cost.
(The realistic $J/U$ values of alkali-doped fullerides are about
a factor of 10 smaller~\cite{PhysRevB.85.155452}.)

\begin{figure}[t]
\begin{center}
\includegraphics[width=8.5cm]{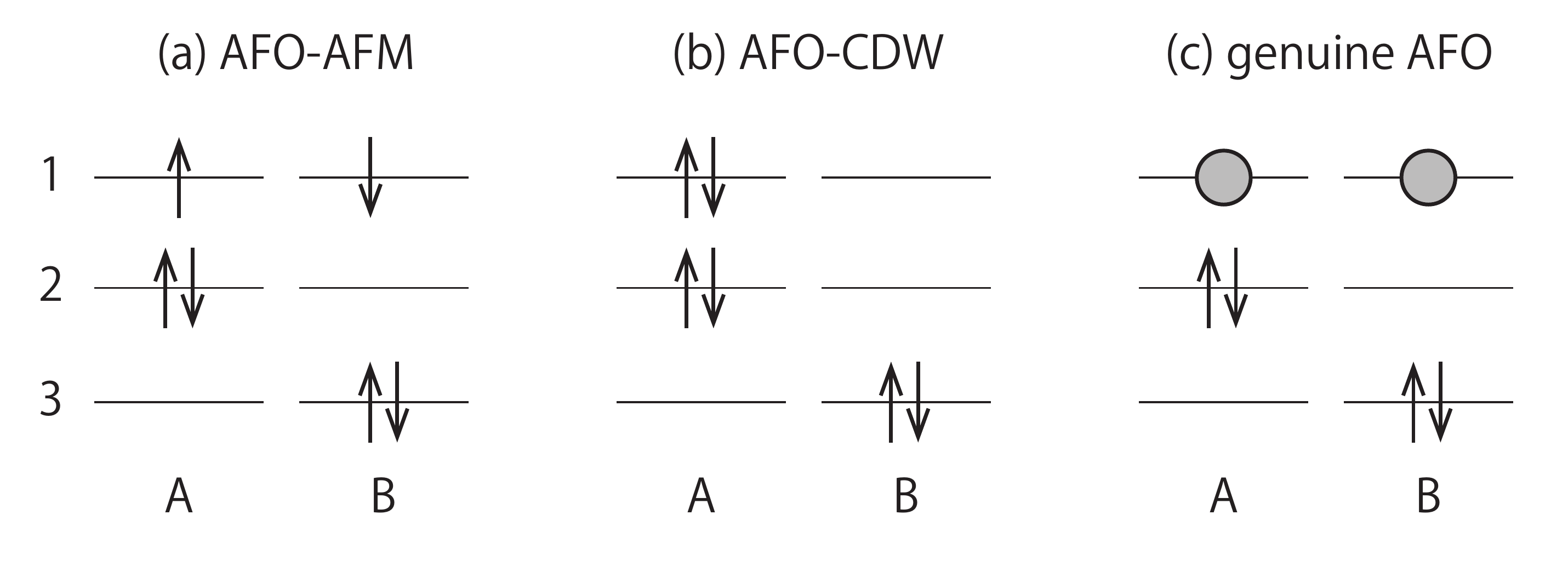}
\caption{
  Schematic pictures for the possible orbital-selective
  staggered ordered states in the three-orbital Hubbard model
  with negative Hund coupling.
  $\alpha(=1,2,3)$ and $\gamma(=A,B)$ represent
  the orbital and sublattice indices, respectively.
}
\label{cdwafoafm}
\end{center}
\end{figure}

We first consider the local electron configurations favored
by ${\cal H}_U$, 
which helps us to discuss the possible ordered states in the three orbital system.
Since the interorbital Coulomb interation $U'(>U)$ is
dominant in the half-filled system, the three orbitals at each site are empty, singly occupied, and doubly occupied, respectively.
This means that there are degrees of freedom for how to distribute these three local states among the oribtals, and the singly occupied orbital has also a spin degree of freedom.
Therefore, in the ground state with negative Hund coupling, it is expected that
the orbital degrees of freedom, in addition to the charge and spin degrees of freedom, will
be ordered. 
The active orbital degrees of freedom are in contrast to the case of positive (or ferromagnetic) Hund coupling, where all three orbitals are singly occupied at half-filling and the orbital degrees of freedom are quenched.
In the strong coupling regime of Eq.~\eqref{3orb}, which will be discussed in detail,
it is naively expected that the exchange coupling between adjacent spins
should induce an AFM order associated with
an AFO state for the empty and doubly occupied orbitals, which is schematically shown in
Fig.~\ref{cdwafoafm}(a).
On the other hand, it is not
obvious whether or not this AFO-AFM state is realized in the weak coupling region.


To clarify this point, let us start with the simple Hartree approximation
with static mean-fields:
$ n_{i\alpha\sigma} n_{i\beta \sigma'}
\longrightarrow
\langle n_{i\alpha\sigma}\rangle n_{i\beta \sigma'}
+ n_{i\alpha\sigma} \langle n_{i\beta \sigma'}\rangle
- \langle n_{i\alpha\sigma}\rangle \langle n_{i\beta \sigma'}\rangle$.
Here, we define the order parameters for the CDW and AFM states, which are given by
\begin{eqnarray}
m_{CDW}&=&\frac{1}{N}\sum_{i\alpha\sigma}(-1)^i
\langle n_{i\alpha\sigma}\rangle,\\
m_{AFM}&=&\frac{1}{N}\sum_{i\alpha\sigma}(-1)^i\sigma
\langle n_{i\alpha\sigma}\rangle.
\end{eqnarray}
On the other hand,
in the three orbital system considered,
possible orbital orders are described by the $3\times 3$ Gell-Mann matrices, which are given by
\begin{align}
\lambda_1=
 \begin{pmatrix}
  0&1&0\\
1&0&0\\
0&0&0
 \end{pmatrix},\quad
\lambda_2=
 \begin{pmatrix}
  0&-i&0\\
i&0&0\\
0&0&0
 \end{pmatrix},\quad
\lambda_3=
 \begin{pmatrix}
  1&0&0\\
0&-1&0\\
0&0&0
 \end{pmatrix},\nonumber\\
\lambda_4=
 \begin{pmatrix}
  0&0&1\\
0&0&0\\
1&0&0
 \end{pmatrix},\quad
\lambda_5=
 \begin{pmatrix}
  0&0&i\\
0&0&0\\
-i&0&0
 \end{pmatrix},\quad
\lambda_6=
 \begin{pmatrix}
  0&0&0\\
0&0&1\\
0&1&0
 \end{pmatrix},\nonumber\\
\lambda_7=
 \begin{pmatrix}
  0&0&0\\
0&0&-i\\
0&i&0
 \end{pmatrix},\quad
\lambda_8=\frac{1}{\sqrt{3}}
 \begin{pmatrix}
  1&0&0\\
0&1&0\\
0&0&-2
 \end{pmatrix}.
\end{align}
Using these matrices, local orbtial moments are given by
\begin{align}
 m^{\rm orb}_{in}=\sum_{\alpha\alpha'\sigma}(\lambda_n)_{\alpha\alpha'} \langle c_{i\alpha\sigma}^\dagger  c_{i\alpha'\sigma}\rangle.
\end{align}
A previous analysis based on generalized susceptibilities~\cite{PhysRevLett.118.177002}  implies that orbital orders with off-diagonal components corresponding to the spontaneous mixing of orbitals are unlikely to appear, and thus
we only consider the two diagonal orbital orders with $\lambda_8$ and $\lambda_3$, which are explicitly given by
\begin{align}
   \xi_{i}&=\frac{1}{2}m^{\rm orb}_{i8}= \frac{ 1}{ 2} \sqrt{\frac{ 1}{ 3}} \sum_{\sigma}
  \left(\langle n_{{i}1\sigma}\rangle+\langle n_{{i} 2\sigma}\rangle-2\langle n_{{i} 3\sigma}\rangle\right),\label{eq:5}\\
 \eta_{i}&=\frac{1}{2}m^{\rm orb}_{i3}= \frac{ 1}{ 2}  \sum_{\sigma} \left(\langle n_{{i}1\sigma}\rangle
 -\langle n_{{i} 2 \sigma}\rangle\right).\label{eq:6}
\end{align}
\begin{figure}[t]
\begin{center}
\includegraphics[width=8cm]{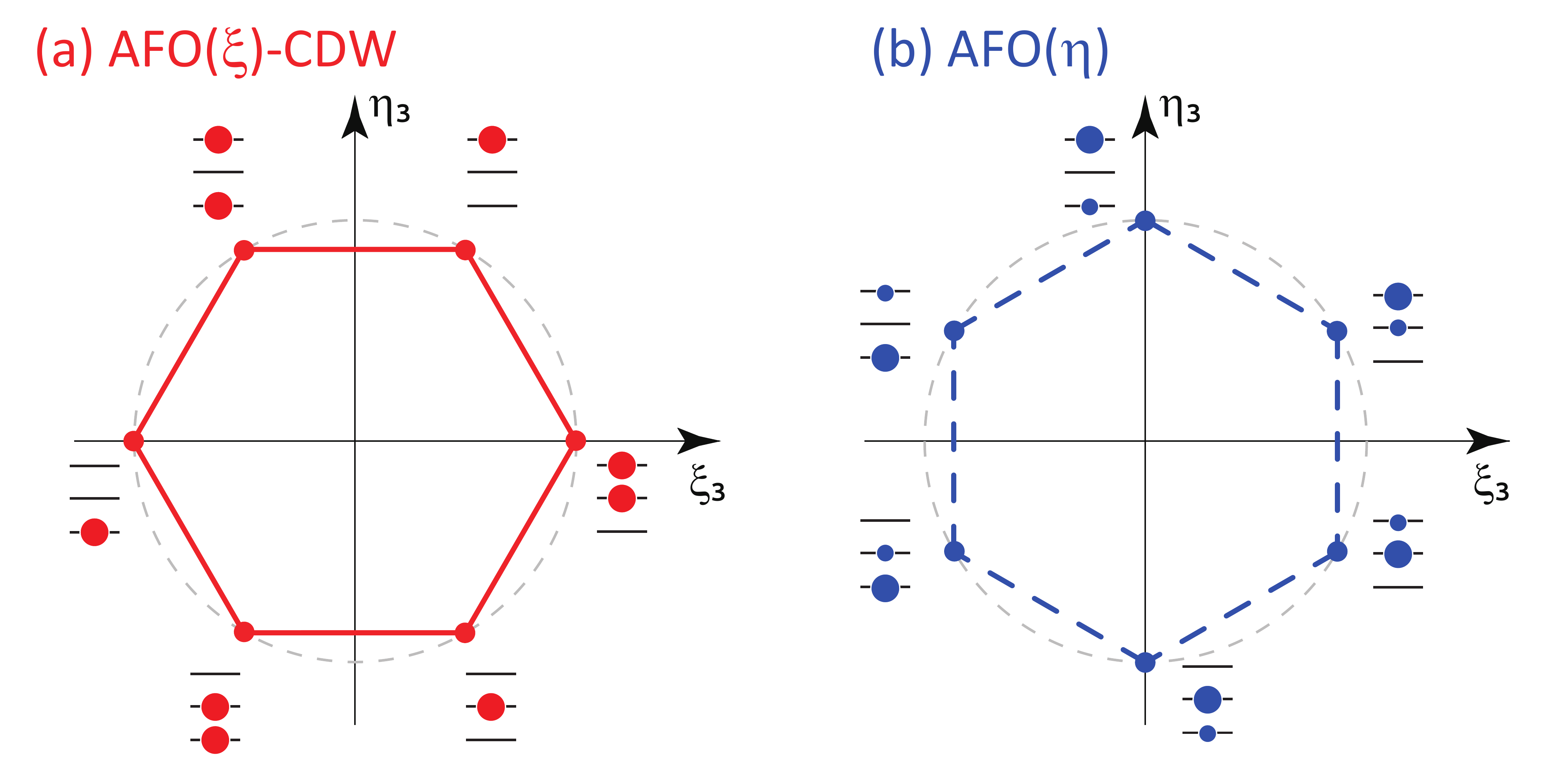}
\caption{
  Classification of orbital orders in the three-orbital Hubbard model.
  The points connected by lines indicate equivalent solutions.
}
\label{fig:tp}
\end{center}
\end{figure}
Note that one can also introduce the symmetrically equivalent orbital moments as follows{}:
\begin{eqnarray}
  \xi_{i\gamma}&=& \frac{ 1}{ 2} \sqrt{\frac{ 1}{ 3}} \sum_{\sigma}
  \left(\langle n_{{i}\alpha\sigma}\rangle+\langle n_{{i} \beta\sigma}\rangle-2\langle n_{{i} \gamma\sigma}\rangle\right),\\
 \eta_{i\gamma}&=& \frac{ 1}{ 2}  \sum_{\sigma} \left(\langle n_{{i}\alpha\sigma}\rangle
 -\langle n_{{i} \beta \sigma}\rangle\right),
\end{eqnarray}
with $(\alpha,\beta,\gamma)=(1,2,3)$ and its cyclic permutations. $\xi_i$ and $\eta_i$ in Eqs.~(\ref{eq:5}) and~(\ref{eq:6}) are then given by $\xi_{i}=\xi_{i3}$ and $\eta_{i}=\eta_{i3}$.
These orbital moments are obtained from $\xi_{i}$ and $\eta_{i}$ by applying the $C_3$ rotation as shown in Fig.~\ref{fig:tp}.
Therefore, we introduce two symmetrically inequivalent AFO order parameters, as
\begin{align}
m_{AFO}^{(\xi)}&=\frac{1}{N}\sum_{i}(-1)^i\sqrt{\xi_i^2+\eta_i^2}\cos\left( 3\arctan \frac{\eta_i}{\xi_i} \right)\nonumber\\
&=\frac{1}{N}\sum_{i}(-1)^i\frac{4\xi_{i1}\xi_{i2}\xi_{i3}}{\xi_{i3}^2+\eta_{i3}^2},\\
m_{AFO}^{(\eta)}&=\frac{1}{N}\sum_{i}(-1)^i\sqrt{\xi_i^2+\eta_i^2}\cos\left( 3\arctan \frac{\xi_i}{\eta_i} \right)\nonumber\\
&=\frac{1}{N}\sum_{i}(-1)^i\frac{4\eta_{i1}\eta_{i2}\eta_{i3}}{\xi_{i3}^2+\eta_{i3}^2}.
\end{align}
These allow us to distinguish two types of AFO states
[AFO($\xi$) and AFO($\eta$)].
For example, one expects $m_{AFO}^{(\xi)}=0$ and $m_{AFO}^{(\eta)}=1$
in the strong coupling limit [see Fig.~\ref{cdwafoafm}(a)].

\begin{figure}[t]
\begin{center}
\includegraphics[width=8cm]{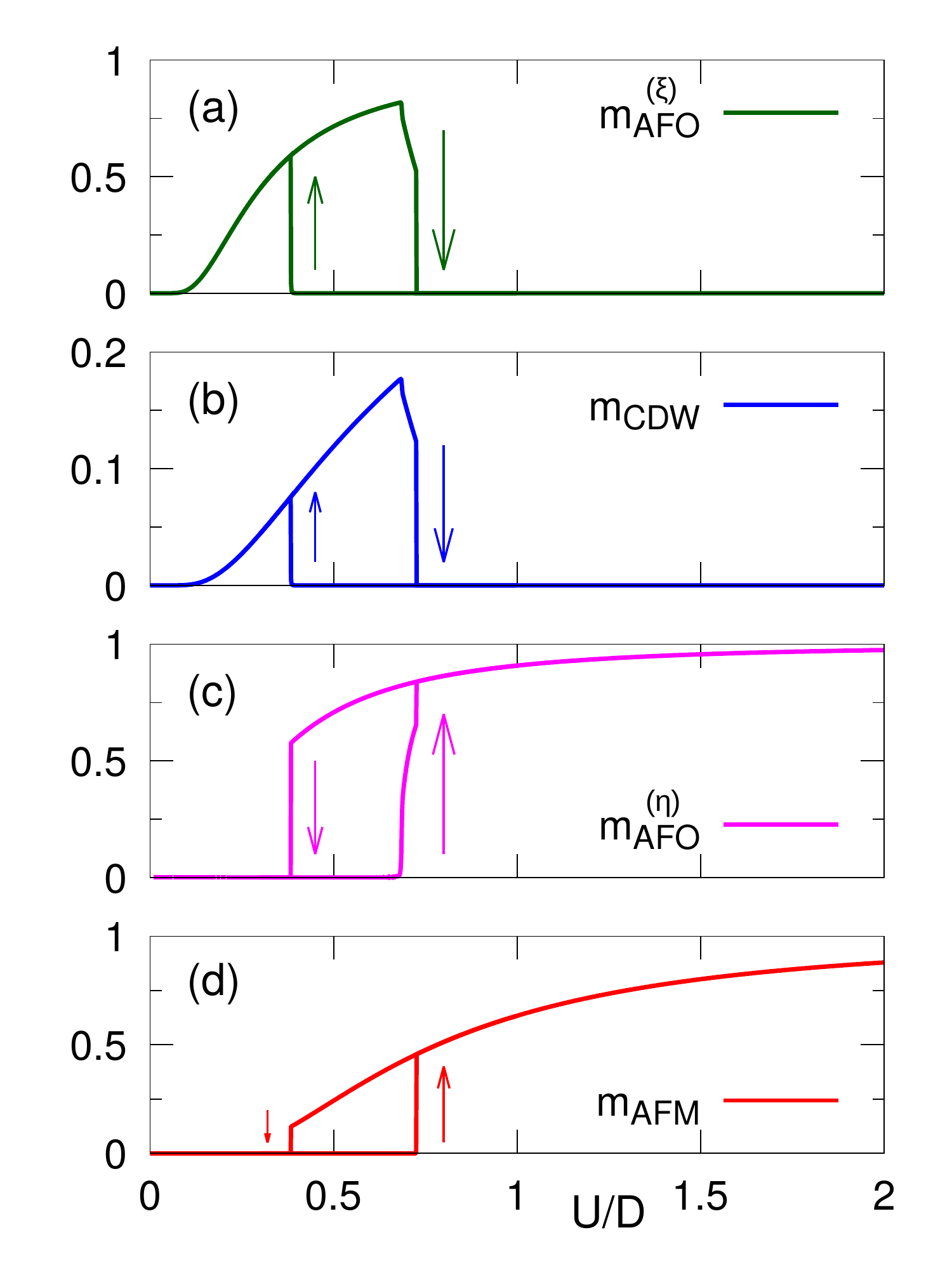}
\caption{Order parameters for the translational symmetry breaking states
obtained from the Hartree approximation.
The arrows indicate the existence of a hysteresis in the order parameters.
}
\label{HF_order}
\end{center}
\end{figure}

\begin{figure}[t]
\begin{center}
\includegraphics[width=8cm]{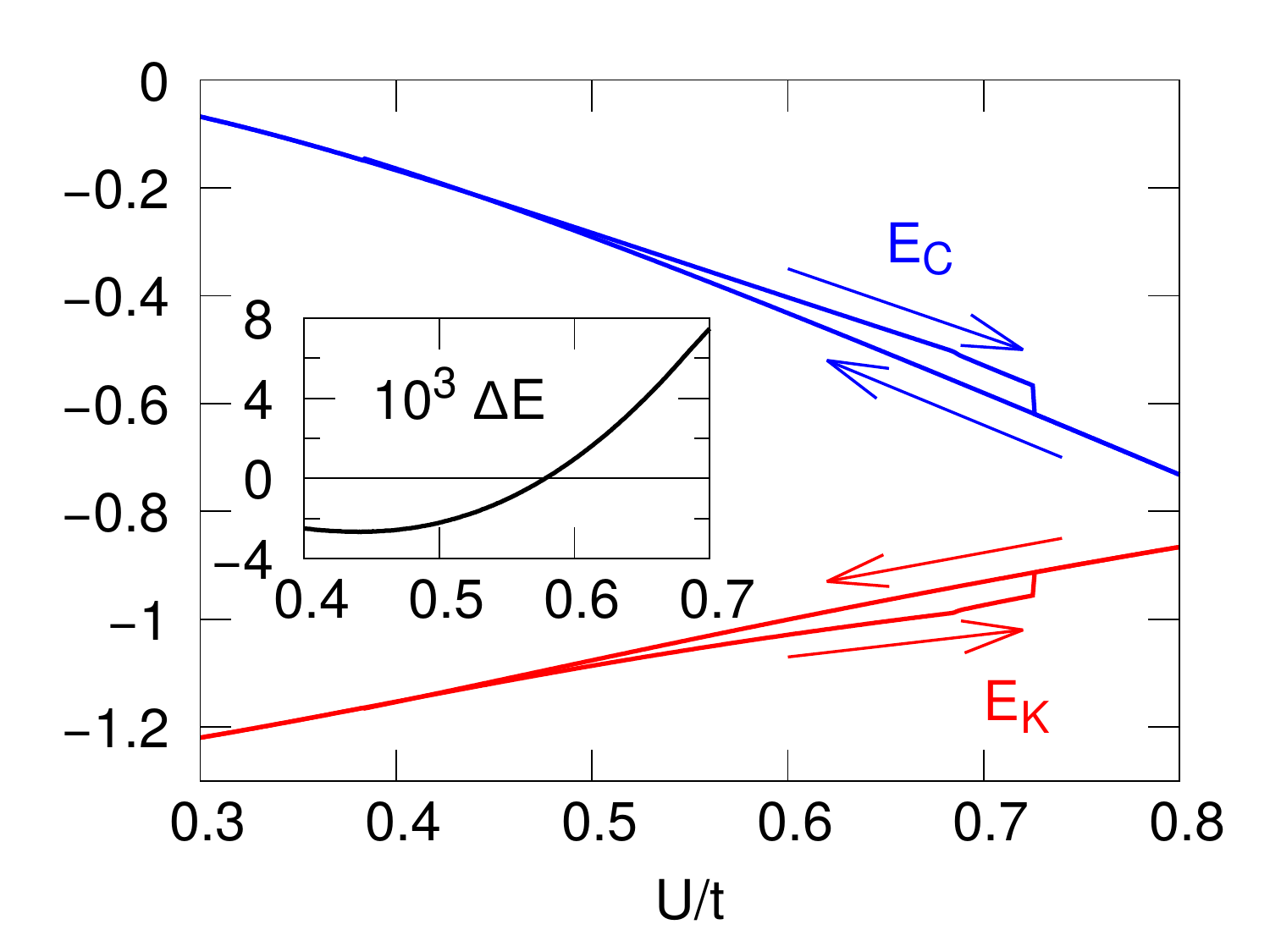}
\caption{
  Kinetic ($E_K$) and correlation ($E_C$) energies as
  a function of $U/t$. The
  inset shows the difference in energies between \AFOAFM and \AFOCDW states.
}
\label{HF_E}
\end{center}
\end{figure}

Figure~\ref{HF_order} shows the results obtained in
the Hartree approximation at zero temperature.
In the strong coupling region, the order parameters for both AFO($\eta$)
and AFM states
are finite, which we refer to as an \AFOAFM state, as expected above.
On the other hand, in the weak coupling limit,
the order parameter for the AFM state disappears, while
that for the CDW state becomes finite.
Note that the different components of the AFO order behave differently.
We find a finite $m_{AFO}^{(\xi)}$, while $m_{AFO}^{(\eta)}$ vanishes.
This suggests the existence of an \AFOCDW state,
which is schematically shown in Fig.~\ref{cdwafoafm}(b).
Around $U/D\sim 0.5$, hystereses regions appear in the order parameters
and a first-order quantum phase transition
takes place between these two states.
This is understood from the absence of the inclusion relation between the \AFOAFM and \AFOCDW states~\cite{Landau_textbook_SP}.
The transition point $(U/D)_c\sim 0.58$ is determined by
the crossing point of two energy curves for \AFOAFM and \AFOCDW states,
as shown in the inset of Fig.~\ref{HF_E}.
Since the kinetic energy gain is larger for the \AFOCDW state
than the \AFOAFM state, see Fig.~\ref{HF_E},
the \AFOCDW state is mainly stabilized by the kinetic energy,
while the \AFOAFM state is stabilized by the correlation energy.
Therefore, the CDW state is realized in the weak coupling region.

These results suggest that the \AFOCDW and \AFOAFM states
compete with each other at zero temperature.
However, it is not clear how well
the Hartree approximation describes
the ground state properties, and in particular
the first-order quantum phase transition in the intermediate coupling
region, since dynamical correlations cannot be taken into account.
Furthermore, it is instructive to study how stable the \AFOAFM and \AFOCDW states are
against thermal fluctuations.
In the following section, we make use of DMFT to discuss the
finite temperature properties. 

\section{DMFT analysis}\label{sec:results}

\subsection{DMFT Framework}

First, we briefly introduce the framework of DMFT.
In DMFT, the lattice model is mapped to an effective impurity model,
which allows to accurately describe local electron correlations.
The lattice Green function is then obtained via a self-consistency conditions
imposed on the impurity problem.
This treatment is exact in infinite dimensions and is expected to
give a qualitatively correct description even in three dimensions.
DMFT has been widely applied to models for strongly correlated electron systems.
The Hubbard model with degenerate orbitals has been extensively discussed
in the framework of DMFT, and interesting phenomena have been revealed, such as
simple
Mott transitions~\cite{Kotliar,Rozenberg,Held,Han,Imai,Koga2band1,Koga2band2,InabaMulti},
orbital-selective Mott transitions~\cite{KogaOSMT,KogaOSMT2,InabaOSMT,InabaOSMT2,MediciOSMT,WernerOSMT,Krylov},
magnetism~\cite{Momoi,Yanatori,Golubeva},
and superconductivity~\cite{PhysRevB.91.085108,Yanatori2,Hoshino2015}.

In the DMFT treatment of the three-orbital Hubbard model,
the lattice Green function matrix is given by
\begin{eqnarray}
  G^{-1}(k,i\omega_n)=G_0(k,i\omega_n)^{-1}-\Sigma(
i\omega_n),
\end{eqnarray}
where $\omega_n=(2n+1)\pi T$ is the Matsubara frequency with integer $n$.
The noninteracting Green function is diagonal
in the spin and orbital spaces, and
\begin{eqnarray}
  \left[G_0(k,i\omega_n)^{-1}\right]_{\alpha\sigma}=i\omega_n+\mu-\epsilon_k,
\end{eqnarray}
where $\epsilon_k$ is the dispersion relation.
In two-sublattice DMFT, the local Green function is given by the
site-dependent local
self-energy $\Sigma_{i\alpha\sigma}(i\omega_n)$ $[=\Sigma^\gamma_{\alpha\sigma}(i\omega_n)]$ as
\begin{eqnarray}
  G^\gamma_{\text{loc}, \alpha\sigma}(i\omega_n)
  &=&\int
  \frac{\rho (x)}{i\omega_n+\mu-x-\Sigma^\gamma_{\alpha\sigma}(i\omega_n)} dx,
\end{eqnarray}
where $\gamma (=\!\!\!\!\! A, B)$ is the sublattice index
and $\rho(x) [=2\sqrt{1-(x/D)^2}/(\pi D)]$
is the density of states of the noninteracting system.
The self-consistency equation
is given by~\cite{Chitra}
\begin{eqnarray}
  {\cal G}_{\alpha\sigma}^\gamma (i\omega_n)=i\omega_n+\mu
  -\frac{D^2}{4}G_{\text{imp},\alpha\sigma}^{\bar \gamma}(i\omega_n),
\label{eq:self}
\end{eqnarray}
where $\bar\gamma$ is the opposite sublattice of $\gamma$.
${\cal G}$ ($G_\text{imp}$) is the
bath
(full) Green's function
of the effective impurity model.
In our study, we use, as an impurity solver,
the hybridization-expansion CTQMC method~\cite{CTQMC,CTQMCREV},
which is a powerful method to
study
finite-temperature properties
regardless of the strength of the interaction.

\subsection{Numerical Results}

\begin{figure}[htb]
 \begin{center}
\includegraphics[width=8cm]{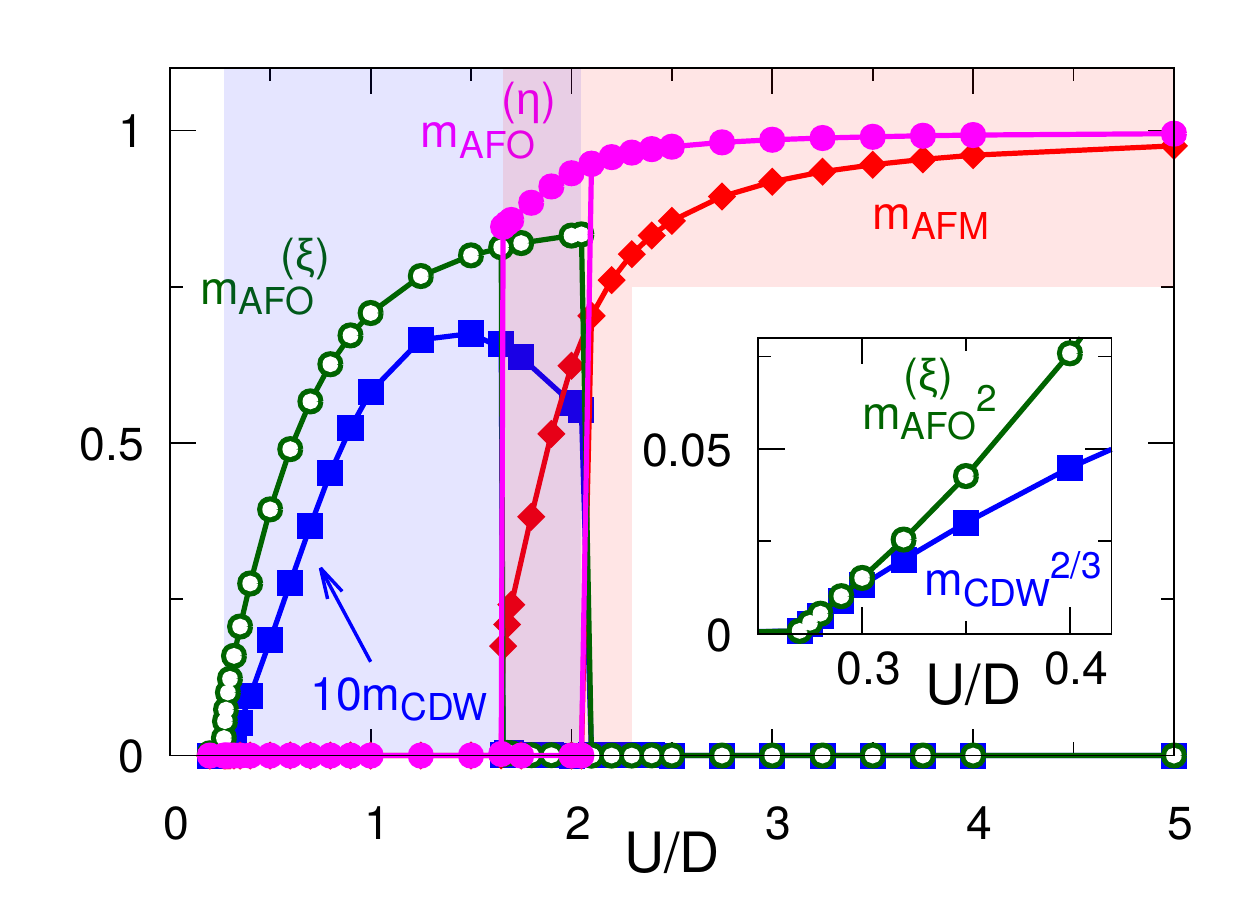}
 \caption{
   Order parameters for the CDW, AFM, and AFO states
   in the bipartite three-orbital system at $T/D=0.01$.
 }
 \label{B100}
 \end{center}
\end{figure}

We now discuss low temperature properties of the three-orbital Hubbard
model with antiferromagnetic Hund coupling.
We iterate the selfconsistency equation [Eq.~\eqref{eq:self}]
to obtain the results within the desired accuracy.
The order parameters at a low temperature $T/D=0.01$
are shown as a function of $U/D$ in Fig.~\ref{B100}.
In the weak coupling region $U<U_c(\sim 0.27D)$,
no order parameters appear and a metallic state is realized.
It is found that, beyond $(U/D)_c$,
nonzero order parameters $m_{AFO}^{(\xi)}$ and $m_{CDW}$ are simultaneoulsly induced.
This shows that the CDW order 
couples to the AFO($\xi$) order, and
the \AFOCDW state is realized in this region.
An interesting point is that the critical behavior of $m_{AFO}^{(\xi)}$ is
different from that of $m_{CDW}$.
We numerically find that the critical exponent of the CDW order parameter takes a nontrivial value $\beta=3/2$, while that of the AFO is $\beta=1/2$,
as shown in the inset of Fig.~\ref{B100}.
These are in good agreement with the results
obtained by the Landau theory,
where the above coupling is taken into account
in the symmetry arguments.
Since $\beta$ should be $1/2$ in the mean-field theory,
the above result clearly indicates that the primary order parameter
is not $m_{CDW}$ but $m_{AFO}^{(\xi)}$.
The details are discussed in Appendix~\ref{sec:Landau}.
Further increase of the interaction increases both order parameters.
Around $U/D\sim 1.5$, the AFO order parameter still increases,
while a nonmonotonic behavior appears in the CDW order parameter.
At last, around $U/D\sim 2$, the \AFOCDW state
suddenly disappears,
and a first-order phase transition to the \AFOAFM state
occurs 
with finite $m_{AFO}^{(\eta)}$ and $m_{AFM}$.
Namely, we find in Fig.~\ref{B100} a solution corresponding to \AFOAFM order
in the region $U/D > 1.66$.
The competition between the \AFOCDW and \AFOAFM states is qualitatively
consistent with the Hartree mean-field results, as discussed in the previous section.

\begin{figure}[htb]
 \begin{center}
 \includegraphics[width=8cm]{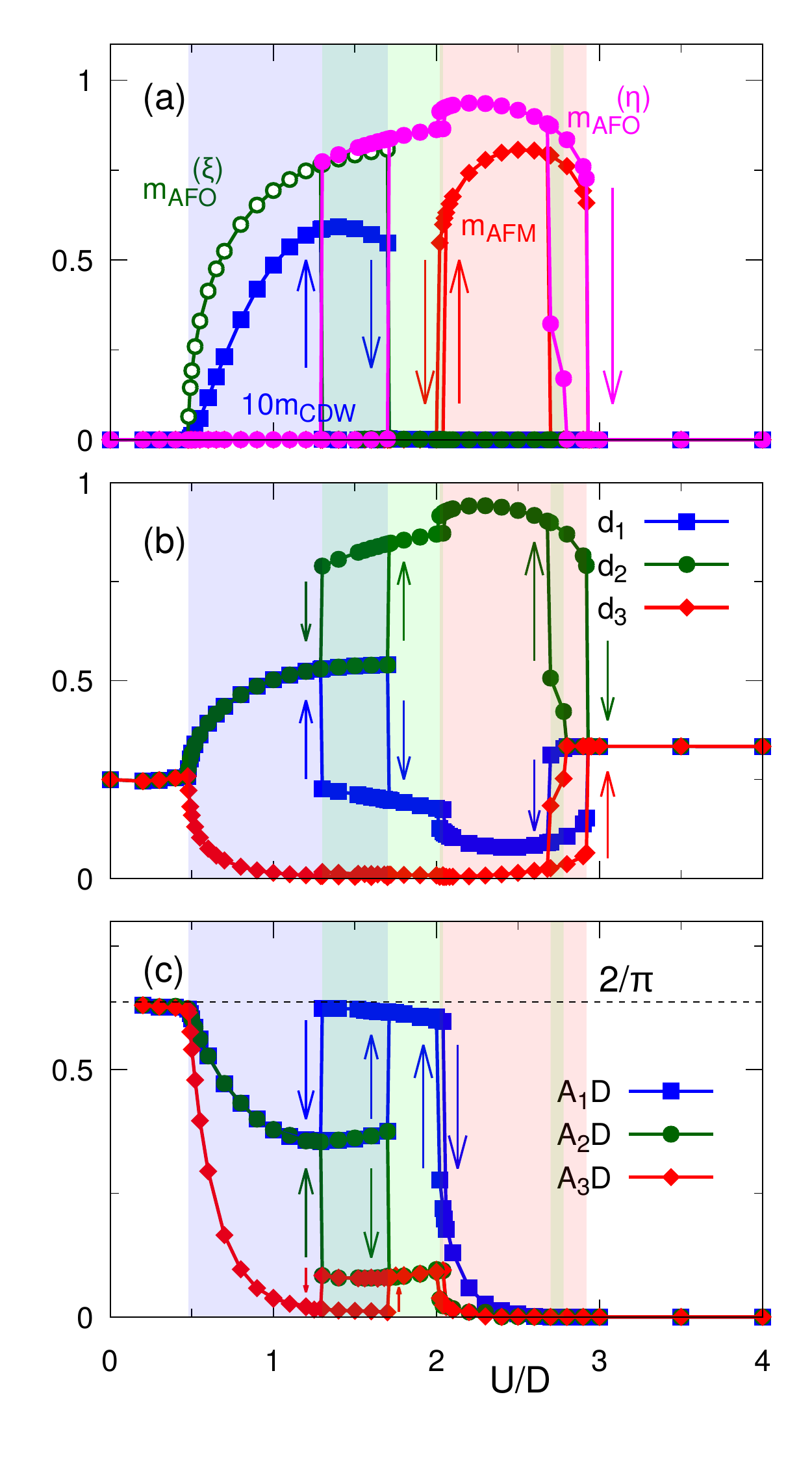}
 \caption{
   (a) Order parameters for the CDW, AFM, and AFO states
   in the bipartite three-orbital system at $T/D=0.04$.
   (b) Double occupancy $d_\alpha$ and (c) the quantity $A_\alpha$
   in one of the sublattices.
 }
 \label{B25}
 \end{center}
\end{figure}

\begin{figure}[htb]
 \begin{center}
 \includegraphics[width=8cm]{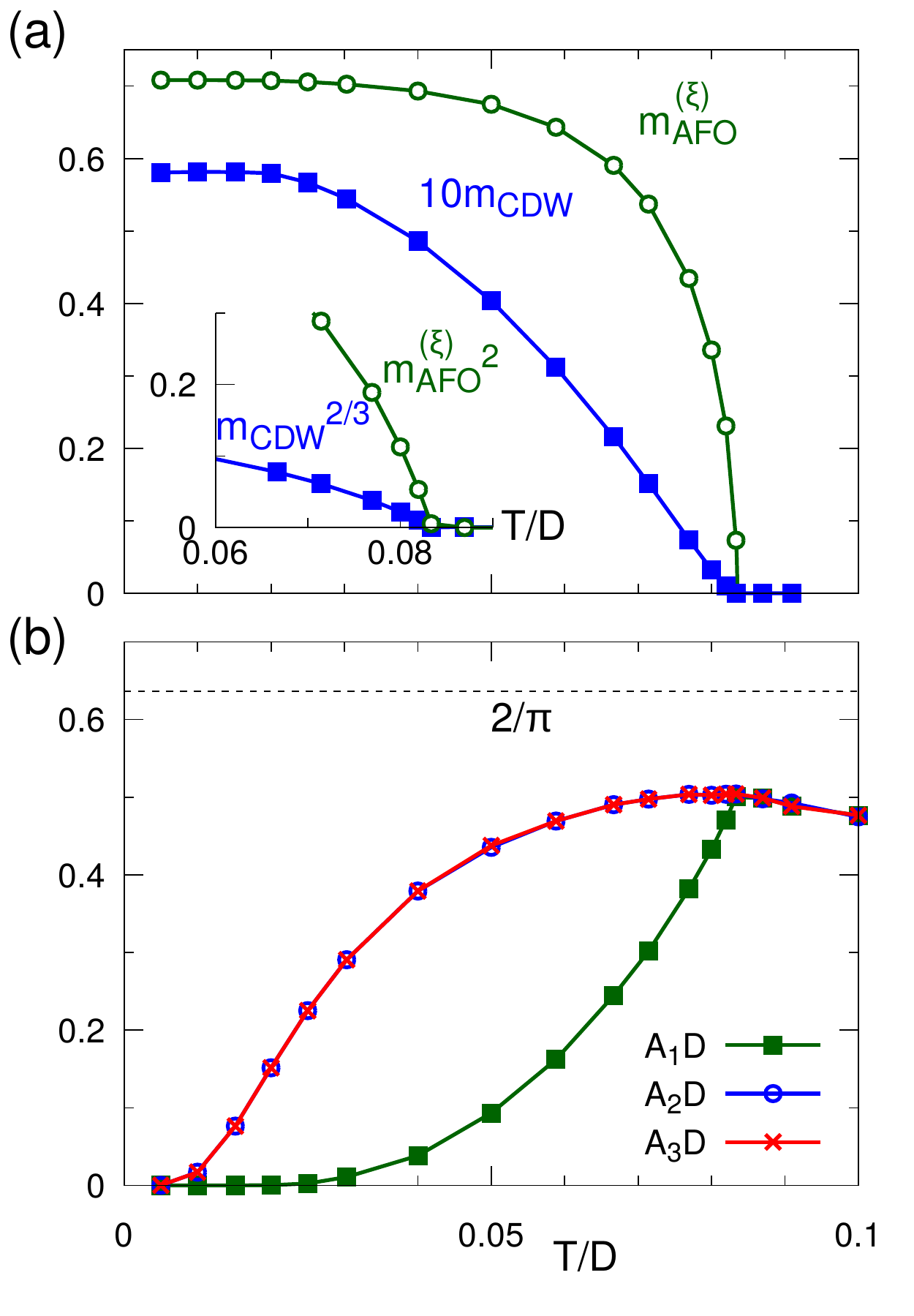}
 \caption{
   Temperature dependence of several quantities
   in the bipartite three-orbital system at $U/D=1.0$.
   (a) Circles and squares represent
   the order parameters $m_{CDW}$ and $m_{AFO}^{(\xi)}$.
   (b) Symbols show the density of states at the Fermi level $A_\alpha$.
 }
 \label{u1}
 \end{center}
\end{figure}

On the other hand, thermal fluctuations
destabilize the \AFOCDW and \AFOAFM states,
which may
result in other staggered ordered states.
Figure~\ref{B25} shows the DMFT results at temperature $T/D=0.04$.
We find that
the \AFOCDW [\AFOAFM] state is realized in the weak (strong) coupling
regions and its solution appears when $0.48<U/D<1.7\;(2.0<U/D<2.9)$.
In addition, there exists
another state around $U/D\sim 2$.
Since $m_{AFO}^{(\eta)}\neq 0$ and $m_{CDW}=m_{AFM}=m_{AFO}^{(\xi)}=0$,
we refer to this state as the \gAFO state.
To clarify the nature of this state,
we also calculate the double occupancy $d_\alpha$ and
the quantity $A_\alpha$ for the $\alpha$th orbital,
which are definied as
\begin{eqnarray}
d_\alpha&=&\langle n_{\alpha\uparrow}n_{\alpha\downarrow}\rangle,\\
A_\alpha&=&-\frac{1}{2\pi}\sum_{n\sigma \gamma}G_{loc,\alpha\sigma}^\gamma(i\omega_n)e^{-i\omega_n/2T}.
\end{eqnarray}
The quantity $A_\alpha$
can be regarded as the density of states at the Fermi level
for the $\alpha$th orbital~\cite{Gull2008,DOST}.
We find that, above the critical interaction $(U/D)_c\sim 0.48$,
the difference between $d_1(=d_2)$ and $d_3$ increases,
and $A_\alpha$ for each orbital rapidly decreases.
To demonstrate the insulating behavior in the \AFOCDW state, 
we examine its temperature dependence.
Figure~\ref{u1} shows that upon 
lowering temperature,
the two order parameters increase while the 
quantities $A_\alpha$ monotonically decrease.
At $T/D\lesssim 0.01$, the densities of states around the Fermi level
almost vanish.
Therefore, an insulating behavior indeed appears in the \AFOCDW state,
which is consistent with the fact that
all orbitals are involved in both the AFO($\xi$) and CDW order parameters.
The \AFOAFM state is also found to be insulating since
the density of states for each orbital is tiny,
as shown in Fig.~\ref{B25}(c).
A qualitatively different behavior appears in the \gAFO phase which is located
between the \AFOCDW and \AFOAFM phases.
A staggered orbital order is realized in two of three orbitals
(orbital 2 and 3) and thereby the double occupancy
takes large and small values in these orbitals
($d_2\sim 0.6$ and $d_3\sim 0.0$).
Since the densities of states at the Fermi level $A_2$ and $A_3$ are smaller than $0.1D^{-1}$,
charge degrees of freedom are almost frozen, and these orbitals are insulating.
On the other hand, the remaining orbital (orbital 1) appears to be metallic as the double occupancy $d_1\sim 0.25$ and the density of states
$A_1D\sim 0.6$.
These results indicate that the \gAFO state remains metallic
even though the neighboring two \AFOCDW and \AFOAFM states are insulating.

\begin{figure}[t]
 \begin{center}
 \includegraphics[width=8cm]{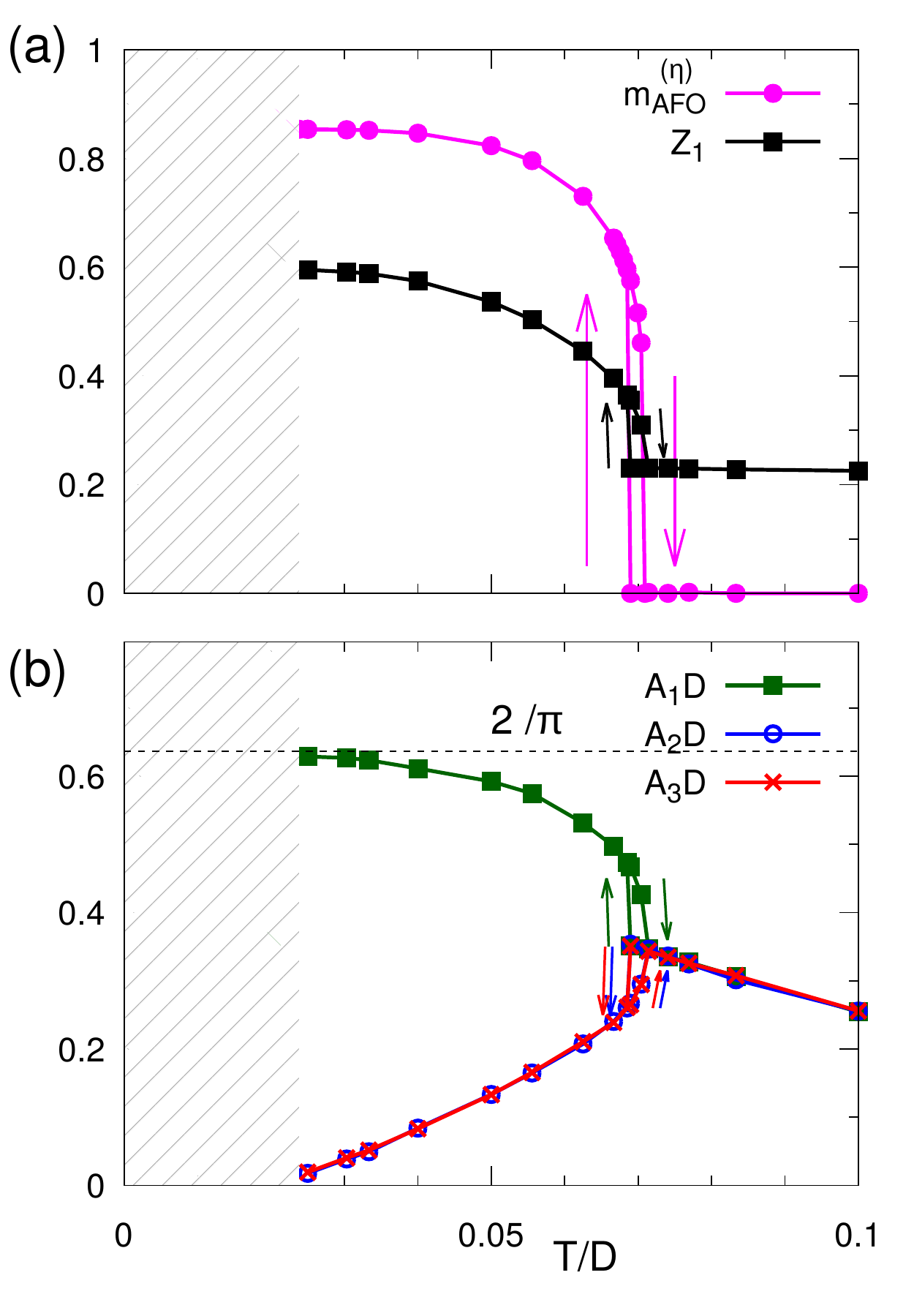}
 \caption{
   Temperature dependence of several quantities
   in the bipartite three-orbital system at $U/D=1.8$.
   (a) Circles and squares represent
   the order parameter $m_{AFO}^{(\eta)}$ and the quantity $Z_1$ for the metallic orbital.
   (b) Symbols show the density of states at the Fermi level $A_\alpha$.
   The shaded area indicates the temperature range where
   the \AFOCDW or \AFOAFM states should be realized
   instead of the \gAFO state.
 }
 \label{U18}
 \end{center}
\end{figure}

To clarify the origin of the 
metallic property of the \gAFO state,
we show in Fig.~\ref{U18} the temperature dependence of
several quantities in the system with $U/D=1.8$.
At intermediate temperatures $(0.03\lesssim T/D \lesssim 0.07)$,
the \gAFO state is realized without $m_{AFM}$ and $m_{CDW}$.
A jump singularity with hysterisis
appears for $m_{AFO}^{(\eta)}$ around $T/D\sim 0.07$,
which implies a first-order transition between
the low-temperature \gAFO and high-temperature paramagnetic phases.
In the AFO state with one metallic and two insulating orbitals,
the renormalization factor $Z_1$ of the metallic orbital increases
with decreasing temperature, as shown in Fig.~\ref{U18}(a).
At the same time, the density of states approaches $2/\pi$,
as shown in Fig.~\ref{U18}(b), indicating metallic behavior in orbital 1 at the lowest temperature.
In this state, orbital 1 is singly occupied whereas orbitals 2 and 3 are empty and doubly occupied.
In the latter two orbitals, charge fluctuations are suppressed due to the associated AFO($\eta$) order.
When one focuses on the orbital 1, the interorbital interactions $U'$ and $J$
are irrelevant since the corresponding interaction energy is not changed. 
Therefore, in this case, only the onsite interaction $U$ is relevant.
However, the interaction $U$ is not large enough to realize a Mott insulating state.
Therefore, in the singly occupied orbital, the quasi-particle peak should develop at relatively high temperatures,
which is consistent with the fact that
the corresponding density of states rapidly approaches $2/\pi$
below the transition temperature $T/D\sim 0.07$.

\begin{figure}[t]
 \begin{center}
 \includegraphics[width=8cm]{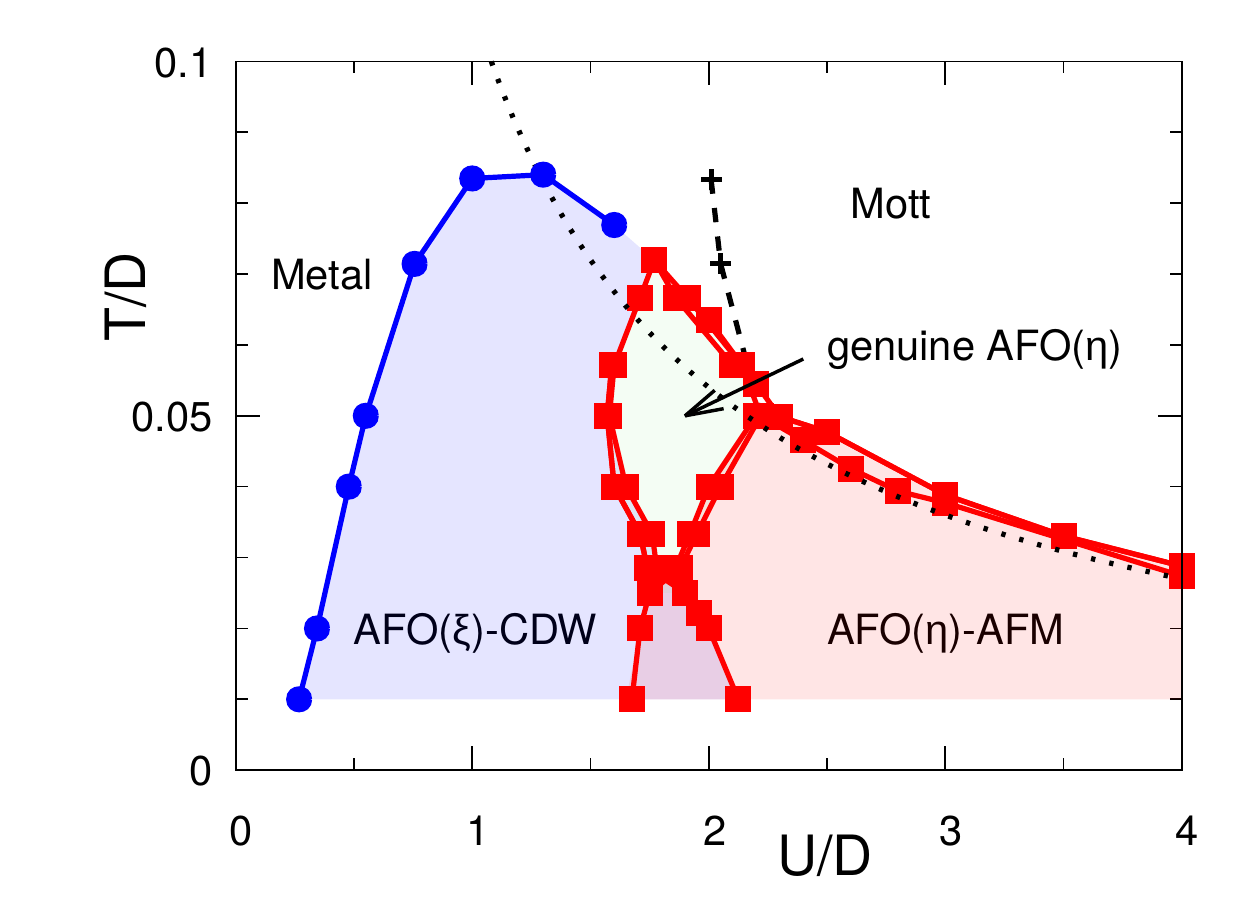}
 \caption{
   Phase diagram for the three-orbital Hubbard model
   on the bipartite lattice.
   The blue line indicates a second order phase transition,
   while the red lines delimit the coexistence regions associated with a first order transition.
   The black dashed line indicates the crossover between metal and Mott insulator
   in the symmetric phase and the dotted line shows the boundary to the AFM state
   predicted by the strong-coupling theory.
 }
 \label{PD}
 \end{center}
\end{figure}

By performing a series of DMFT calculations,
we obtain the finite temperature phase diagram shown in Fig.~\ref{PD}.
We find that the CDW state appears always together with the AFO($\xi$) state.
The phase transition between the \AFOCDW and paramagnetic states
is of second order in the weak coupling region.
The critical phenomena will be discussed in Appendix~\ref{sec:Landau}.
Around $U/D\sim 1.5$, the nature of the phase transition changes to first order.
It is also found that the \gAFO state is realized
only at finite temperatures.
This originates from the fact that
the metallic orbital gains an entropy $S\sim \gamma T$
at nonzero temperature,
where $\gamma$ is the specific heat coefficient.
The phase transition between the \gAFO and paramagnetic states
is of first order.
The \AFOAFM state is realized in the strong coupling region,
as expected from the Hartree approximation.
The phase diagram in Fig.~\ref{PD} has a triple point
at $(U/D,T/D) \sim (1.8,0.027)$ where three first-order lines terminate.
The nature of this point is discussed in more detail in Appendix B.

To investigate the stability of the metallic state,
we have evaluated the Mott transition point by constraining the solution to paraorbital and paramagnetic states.
There exists a Mott critical end point around
$(U/D,T/D)\sim (2.3, 0.04)$.
However, this is located in the \AFOAFM phase in our phase diagram,
and hence the Mott transition never occurs
in our bipartite system.
The Mott crossover line, which is determined by the inflection point of
the curve of the renormalization factor, appears at higher temperatures,
which is shown as the dashed line in Fig.~\ref{PD}.
We find that the \gAFO state is located close to the Mott crossover line,
in other words, 
the \gAFO state with one metallic orbital is stable rather than
the strongly correlated metallic state with equivalent orbitals, and it is realized
between the \AFOCDW and \AFOAFM insulating states.
This is characteristic of the three-orbital model with antiferromagnetic
Hund coupling.

As for the strong coupling region,
the phase transition between the \AFOAFM and Mott states should be first order
although a conventional symmetry breaking occurs in the spin sector.
In the next section,
we use the strong coupling theory
to discuss the phase transition between these states.

\section{Strong coupling theory}\label{sec:strong}

\begin{figure*}[tb]
\begin{center}
\includegraphics[width=2\columnwidth,clip]{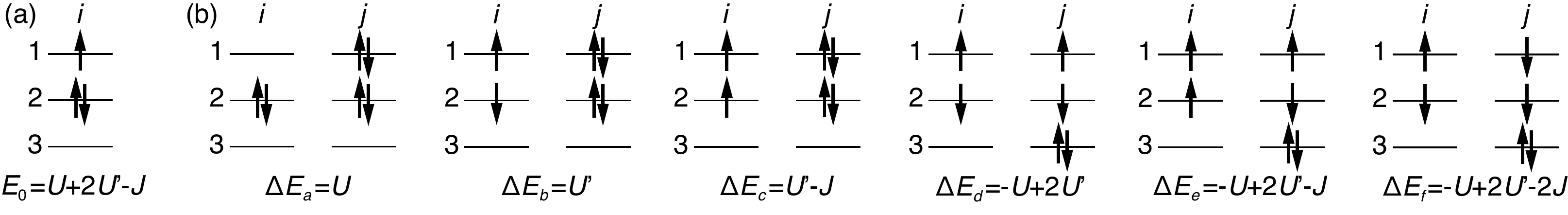}
\caption{(a) One of the localized states in the model space with energy $E_0$ in the strong coupling limit and (b) representative configurations of intermediate states with six different energies.
}
\label{intermediate}
\end{center}
\end{figure*}
\begin{figure*}[tb]
\begin{center}
\includegraphics[width=1.5\columnwidth,clip]{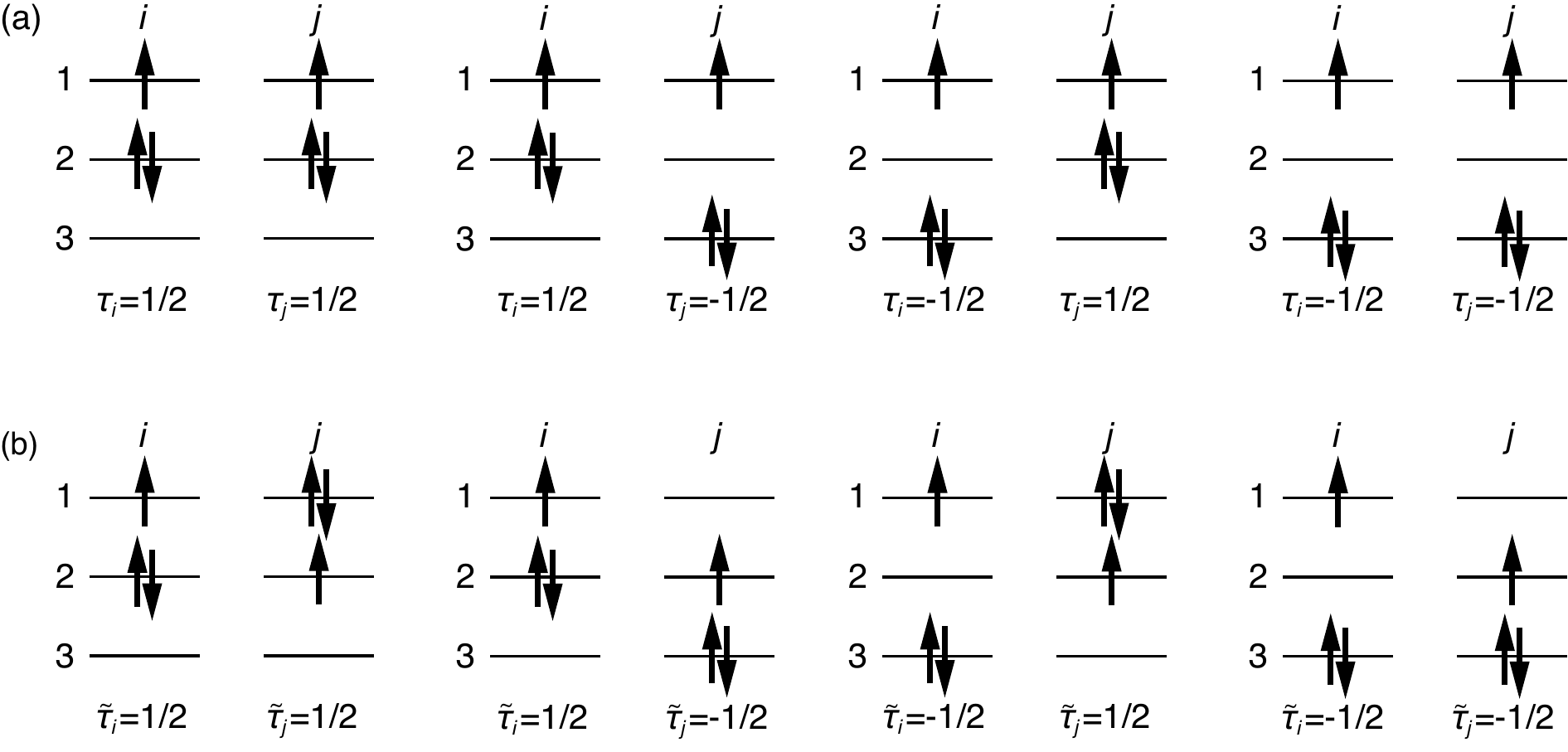}
\caption{Orbital pseudospin configurations for nearest-neighbor states where (a) the singly-occupied orbitals are the same and (b) where they are different.
}
\label{exchange_config}
\end{center}
\end{figure*}

In this section, we discuss the finite temperature
phase transitions based on the strong coupling theory.

\subsection{Effective Hamiltonian}

First, we derive the effective Hamiltonian
in the strong coupling limit.
In the atomic limit with $U<U'$ and $J<0$,
there exist 12 dominant local states with three electrons
and eigenenergy $E_0=U+2U'-J$,
as shown in Fig.~\ref{intermediate}(a).
The superexchange Hamiltonian can be written as
\begin{align}
 {\cal H}_{\rm eff}=-\sum_{\means{ij}}\sum_{\alpha\alpha'} {\cal P}_{i}^{(\alpha)}{\cal P}_{j}^{(\alpha')}{\cal H}_t^{ij} \frac{1}{{\cal H}_U-E_0N}{\cal H}_t^{ij}{\cal P}_{i}^{(\alpha)}{\cal P}_{j}^{(\alpha')},\label{eq:1}
\end{align}
where $N$ is the number of sites, ${\cal P}_{i}^{(\alpha)}$ is the projection operator at site $i$ onto those local states among the 12 configurations introduced above for which the $\alpha$th orbital is singly occupied and ${\cal H}_t^{ij}$ represents the electron transfer between sites $i$ and $j$ in ${\cal H}_t$.
Here, we neglect the exchange process for singly-occupied orbitals on neighboring sites, corresponding to off-diagonal contributions with respect to $\alpha$, as it does not affect the mean-field results, which is discussed later.

Figure~\ref{intermediate}(b) shows the representative six eigenstates
of ${\cal H}_U$ which we consider as the intermediate states.
The energy differences $\Delta E_a,\cdots,\Delta E_f$ from the ground state
corresponding to the denominator of Eq.~(\ref{eq:1}) are also given in this figure.
In the following, we discuss the superexchange processes in two parts;
the effective Hamiltonian is divided as
\begin{align}
 {\cal H}_{\rm eff}={\cal H}_{\rm eff 1}+{\cal H}_{\rm eff 2},
\end{align}
where the first term corresponds to the case with $\alpha=\alpha'$ and the second term to $\alpha\neq \alpha'$.

First, we focus on the case with $\alpha=\alpha'$ in Eq.~(\ref{eq:1}), i.e., the case where the singly-occupied orbitals on  neighboring sites are the same.
In this case, the remaining two are either empty or doubly-occupied orbitals.
To characterize the orbital configuration, we introduce the orbital pseudospin operator $\bm{\tau}_i$ at the $i$th site so that $\tau^z_i=1/2$ ($-1/2$)
when the $\beta$ ($\gamma$) orbital is doubly occupied.
Here, $\alpha'$ and $\alpha''$ are determined depending on $\alpha$ as
$(\alpha,\beta,\gamma)=(1,2,3)$ and its cyclic permutations [see Fig.~\ref{exchange_config}(a)].
Note that the exchange process between empty and doubly-occupied orbitals does not exist within the second order perturbation.
Thus, the superexchange terms including $\tau^x$ and $\tau^y$ do not appear, and we may simply write $\tau_i^z=\tau_i$.
In addition, there is a spin degree of freedom on the singly occupied orbital at each site,
which is denoted by $\bm{S}_i$.

Using these operators, the superexchange Hamiltonian is given by
\begin{align}
 {\cal H}_{\rm eff 1}=\sum_{\means{ij}}\sum_\alpha{\cal P}_{i}^{(\alpha)}{\cal P}_{j}^{(\alpha)}\Bigl(C_1+ J_{\tau 1} \tau_i \tau_j +J_{ss 1} \bm{S}_i\cdot \bm{S}_j\notag\\
+J_{s 1} S^z_i S^z_j
 +J_{s\tau 1} S^z_i S^z_j \tau_i \tau_j \Bigr) {\cal P}_{i}^{(\alpha)}{\cal P}_{j}^{(\alpha)},\label{eq:2}
\end{align}
where $C_1=-t^2\Big(\frac{1}{\Delta E_a}+\frac{1}{2\Delta E_d}+\frac{1}{\Delta E_e}+\frac{1}{2\Delta E_f}\Big)$, $J_{\tau 1}=2t^2\Big(\frac{1}{\Delta E_d}+\frac{2}{\Delta E_e}+\frac{1}{\Delta E_f}\Big)$, $J_{ss1}=\frac{4t^2}{\Delta E_a}$, $J_{s1}=2t^2\Big(\frac{1}{\Delta E_d}-\frac{2}{\Delta E_e}+\frac{1}{\Delta E_f}\Big)$, and $J_{s\tau 1}=-4J_{s1}$.
Note that the effective Hamiltonian does not have the SU(2) symmetry in the spin space
since we have neglected the spin exchange in the Hund coupling in the original three-orbital Hubbard model.

Next, we discuss the case with $\alpha\neq \alpha'$.
As in the previous case, there is the spin degree of freedom $\bm{S}_i$ in the singly-occupied orbital and the orbital degree of freedom specifying the doubly-occupied and empty orbitals.
Here, we introduce another orbital pseudospin $\tilde{\tau}_i$ at site $i$, which is defined to be $+1/2$ ($-1/2$) when the orbital which is doubly-occupied at site $i$ is singly-occupied (empty) at the interacting nearest-neighbor site $j$ [see Fig.~\ref{exchange_config}(b)].
Performing the perturbation expansion, we obtain the superexchange Hamiltonian
\begin{align}
 {\cal H}_{\rm eff 2}=\sum_{\means{ij}}\sum_{\alpha\neq \alpha'}{\cal P}_{i}^{(\alpha)}{\cal P}_{j}^{(\alpha')}\Bigl(C_2+ J_{\tau 2} \tilde{\tau}_i \tilde{\tau}_j +J_{s 2} S^z_i S^z_j\notag\\
 +J_{s\tau 2} S^z_i S^z_j \tilde{\tau}_i \tilde{\tau}_j \Bigr) {\cal P}_{i}^{(\alpha)}{\cal P}_{j}^{(\alpha')},\label{eq:3}
\end{align}
where $C_2=-t^2\Big(\frac{1}{\Delta E_b}+\frac{1}{\Delta E_c}+\frac{1}{4\Delta E_d}+\frac{1}{2\Delta E_e}+\frac{1}{4\Delta E_f}\Big)$, $J_{\tau 2}=t^2\Big(\frac{1}{\Delta E_d}+\frac{2}{\Delta E_e}+\frac{1}{\Delta E_f}\Big)$, $J_{s 2}=t^2\Big(\frac{4}{\Delta E_b}-\frac{4}{\Delta E_c}+\frac{1}{\Delta E_d}-\frac{2}{\Delta E_e}+\frac{1}{\Delta E_f}\Big)$, and $J_{s\tau 2}=-4t^2\Big(\frac{1}{\Delta E_d}-\frac{2}{\Delta E_e}+\frac{1}{\Delta E_f}\Big)$.

Combining the two terms given in Eqs.~(\ref{eq:2}) and~(\ref{eq:3}), we obtain the total effective Hamiltonian as follows:
\begin{widetext}
\begin{align}
 {\cal H}_{\rm eff}=J_p\sum_{\means{ij}\alpha}{\cal P}_{i}^{(\alpha)}{\cal P}_{j}^{(\alpha)}
+ \sum_{\means{ij}\alpha}\Bigl( J_{\tau 1} \tau_i^{(\alpha)} \tau_j^{(\alpha)} +J_{ss 1} \bm{S}_i^{(\alpha)}\cdot \bm{S}_j^{(\alpha)}+J_{s 1} S_i^{z(\alpha)} S_j^{z(\alpha)}
 +J_{s\tau 1} S_i^{z(\alpha)} S_j^{z(\alpha)} \tau_i^{(\alpha)} \tau_j^{(\alpha)} \Bigr)\notag\\
+\sum_{\means{ij}\alpha\neq \alpha'}\Bigl(J_{\tau 2} \tilde{\tau}_i^{(\alpha)} \tilde{\tau}_j^{(\alpha')} +J_{s 2} S_i^{z(\alpha)} S_j^{z(\alpha')}
 +J_{s\tau 2} S_i^{z(\alpha)} S_j^{z(\alpha')} \tilde{\tau}_i^{(\alpha)} \tilde{\tau}_j^{(\alpha')} \Bigr)+{\rm const.},\label{eq:4}
\end{align}
\end{widetext}
where we used the relation $\sum_\alpha {\cal P}_{i}^{(\alpha)}=1$ and $J_p=C_1-C_2$.
We furthermore introduced $\bm{S}_i^{(\alpha)}={\cal P}_{i}^{(\alpha)}\bm{S}_i{\cal P}_{i}^{(\alpha)}$,
$\tau_i^{(\alpha)}={\cal P}_{i}^{(\alpha)}\tau_i{\cal P}_{i}^{(\alpha)}$,
and $\tilde{\tau}_i^{(\alpha)}={\cal P}_{i}^{(\alpha)}\tilde{\tau}_i{\cal P}_{i}^{(\alpha)}$.

\subsection{Mean-field approximation}

Here, we apply the mean-field approximation to Eq.~(\ref{eq:4}).
In the case of $J<0$, $J_p$ is always negative, leading to a state with a uniform $\means{{\cal P}^{(\alpha)}}$.
Thus, the spin and orbital configurations are dominated by ${\cal H}_{\rm eff1}$, where $J_{\tau 1}$, $J_{s1}$, and $J_{ss1}$ are positive.
This is consistent with the result of the AFM phase with the AFO$(\eta)$ order in Fig.~\ref{PD}.

Before discussing the bipartite case, we briefly consider the case with translational symmetry in the strong coupling limit.
We assume that $\means{{\cal P}^{(\alpha)}}$ are finite and other mean fields vanish.
The parameter $x$ is introduced as $\means{{\cal P}^{(1)}}=x$ and $\means{{\cal P}^{(2)}}=\means{{\cal P}^{(3)}}=(1-x)/2$ so as to satisfy the constraint of the projection operators.
We expect that below a critical temperature, the mean-field solution $x$ deviates from $1/3$.
The deviation corresponds to the transition to the SOSM-1 phase from the Mott phase.
The mean-field free energy is given by
\begin{align}
 F_{\rm MF}=\frac{z|J_p|}{2}\left(\frac{3}{2}x^2-x+\frac{1}{2}\right)-\frac{1}{\beta}\ln\left(e^{\beta z|J_p| x}+2e^{\beta z|J_p| (1-x)/2}\right),
\end{align}
where $z$ is the coordination number.
From this representation, we find that a first-order transition occurs at $T_c=z|J_p|/(4\ln 2)\simeq 0.361zJ_p$.
At this point, $F_{\rm MF}$ has two minima at $x=1/3$ and $2/3$ giving the same value.
The smaller-$x$ solution exists above $T_1=z|J_p|/3$ and is fixed to $x=1/3$.
On the other hand, the larger-$x$ solution increases and approaches $x=1$ with decreasing temperature and survives below $T_2\simeq 0.364z|J_p|$, which is determined from the following relations:
\begin{align}
 \frac{1-3x'}{3x'(1-x')}=\ln\frac{1-x'}{2x'},\quad T_2=\frac{2z|J_p|}{3x'(1-x')}.
\end{align}
These indicate the existence of the hysteresis region in $T_1<T <T_2$ associated with the first order phase transition at $T_c$.
Let us compare the present result with the phase diagram obtained by DMFT at $J=-U/4$ with $U'=U-2J$.
The strong coupling theory suggests $T_c=\frac{527}{40320\ln 2}\frac{D^2}{U}\simeq 0.0189\frac{D^2}{U}$, where we replace $4zt^2$ by $D^2$ to take into account the Bethe lattice.
This agrees well with the boundary between the Mott and SOSM-1 phases shown in Refs.~\cite{PhysRevLett.118.177002,Ishigaki}.

Next, we consider the bipartite case.
To describe the AFM state appearing in the strong coupling regime of the three-orbital Hubbard model shown in Fig.~\ref{PD}, the following two-sublattice mean fields are assumed:
$\means{{\cal P}^{(1)}}_A=\means{{\cal P}^{(1)}}_B\neq 0$, $\means{{\cal P}^{(2)}}_A=\means{{\cal P}^{(2)}}_B=\means{{\cal P}^{(3)}}_A=\means{{\cal P}^{(3)}}_B\neq 0$, $\means{S^{z(1)}}_A=-\means{S^{z(1)}}_B\neq 0$, $\means{\tau^{(1)}}_A=-\means{\tau^{(1)}}_B\neq 0$, $\means{S^{z(1)}\tau^{(1)}}_A=\means{S^{z(1)}\tau^{(1)}}_B\neq 0$, and the other mean fields are zero, where $A$ and $B$ are the suffix identifying the sublattice.
The mean fields are computed by solving self-consistent equations iteratively.

\begin{figure}[b]
\begin{center}
\includegraphics[width=0.75\columnwidth,clip]{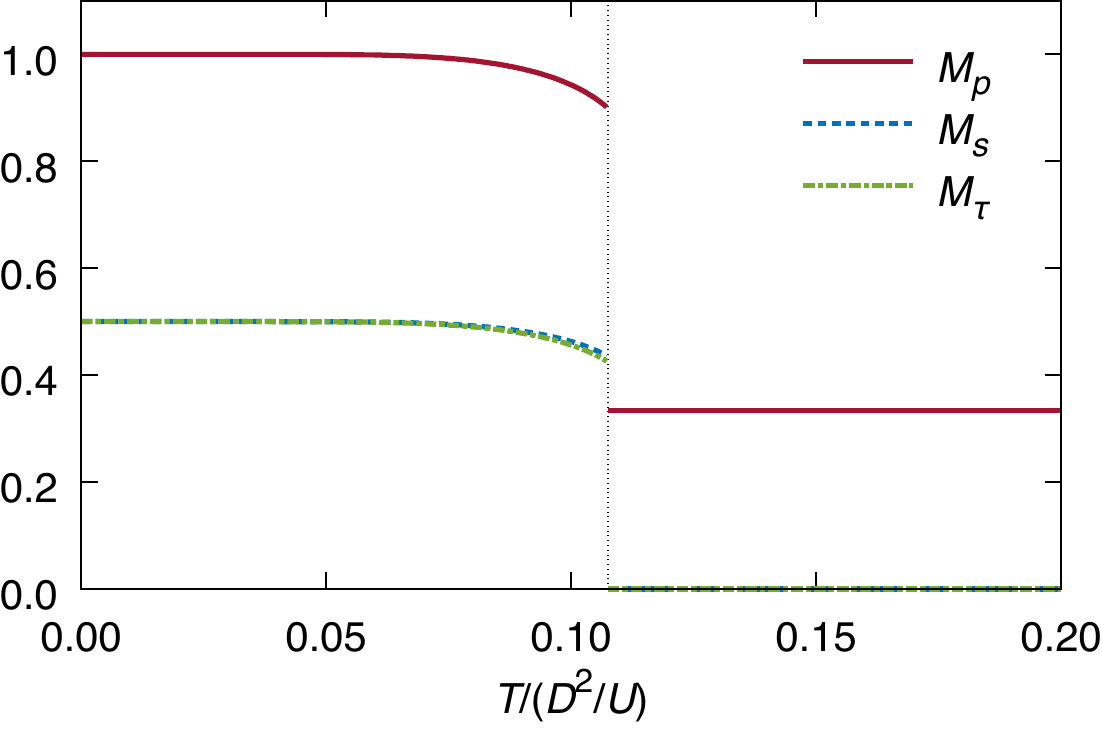}
\caption{Temperature dependence of order parameters $M_p=\means{{\cal P}}$, $M_s=\means{S^z}$, and $M_\tau=\means{\tau}$.
}
\label{strong_Tc}
\end{center}
\end{figure}

Figure~\ref{strong_Tc} shows the temperature dependence of the mean fields at $J=-U/4$ with $U'=U-2J$.
As shown in this figure, a first order phase transition appears at $T_c/D \simeq 0.108 D/U$.
This result is indicated in Fig.~\ref{PD} by the dotted line.
It agrees well with the order of the transition and
its the boundary between the Mott and AFM phases obtained by DMFT.

\section{Summary}
We have considered the half-filled three-orbital Hubbard model
with antiferromagnetic Hund coupling,
combining DMFT with a numerically exact CTQMC impurity solver.
By calculating the electron occupancy in each spin, orbital and sublattice,
we have studied the stability of the AFO, AFM, and CDW states.
We showed that the AFM and CDW states appear
simultaneously with two types of AFO orders,
and these \AFOAFM and \AFOCDW states are separated by
a first order quantum phase transition.
The \AFOAFM state is well described by the superexchange interaction between the nearest neighbor sites in the strong coupling limit.
We have also clarified that a metallic \gAFO state exists
between the two insulating \AFOCDW and \AFOAFM states.
The nontrivial $\beta=3/2$ exponent for the CDW state,
which derives from a characteristic property of the three orbital model,
has also been discussed in terms of the Landau theory.

\begin{acknowledgments}
Parts of the numerical calculations were performed
in the supercomputing systems in ISSP, the University of Tokyo.
This work was supported by Grant-in-Aid for Scientific Research from
JSPS, KAKENHI Grant Nos. JP18K04678, JP17K05536 (A.K.),
JP16K17747, JP16H02206, JP18H04223 (J.N.),
JP16H04021, JP18K13490 (S.H.)
and the European Research Council through ERC Consolidator Grant 724103 (P.W.).
The simulations have been performed using some of
the ALPS libraries~\cite{alps2}.

\end{acknowledgments}

\appendix

\section{Landau theory}\label{sec:Landau}

We use Landau theory based on symmetry arguments to discuss
critical phenomena for the CDW state.
The order parameters considered here are
the conventional staggered orbital and charge moments,
which are defined as
\begin{align}
\xi &= \frac{ 1}{ 2} \sqrt{\frac{ 1}{ 3}} \sum_{i\sigma} (-1)^{i} \left(n_{{i} 1\sigma}+n_{{i} 2\sigma}-2n_{{i} 3\sigma}\right),
\\
\eta &= \frac {1}{ 2} \sum_{i\sigma} (-1)^{i} \left(n_{{i} 1\sigma}-n_{{i} 2\sigma}\right),
\\
\zeta &=\frac{ 1}{ 2} \sum_{i\sigma} (-1)^{i}
 \left(n_{{i} 1\sigma}+n_{{i} 2\sigma}+n_{{i} 3\sigma}\right).
\end{align}
An important point in the bipartite system is that,
in addition to the permutation symmetry in the orbital space,
the system has an exchange symmetry in the sublattice indices $A$ and $B$.
Namely, the order parameters $(\xi,\eta,\zeta)$, which are odd under inversion, must enter in the Landau free energy in products with an even number of factors.
Note that a simple third order term $\xi(\xi^2-3\eta^2)$ does not exist
in the free energy since it is also odd under the exchange.
This is in contrast to the spontaneously orbital-selective Mott cases
discussed in our previous work~\cite{Ishigaki}.
The free energy should be expanded up to the fourth order as
\begin{align}
F = F_0 + a(\xi^2+\eta^2) + a'\zeta^2 + b \zeta \xi(\xi^2 - 3\eta^2) + c(\xi^2+\eta^2)^2.\label{F}
\end{align}
Here, we have omitted the fourth order term $\zeta^4$
since the constant $a'$ should be always positive.
This has been confirmed by the
DMFT calculculation for the paramagnetic and paraorbital states,
where the charge susceptibiltiy never diverges.
For this reason, the genuine CDW state with $\xi=\eta=0$ and $\zeta\neq 0$
is never stabilized.
This allows us to restrict our discussions to the $\xi-\eta$ plane.

When we focus on the instability for the orbital ordered state,
the parameters can be fixed as $a \propto t - t_c$ and $c>0$,
where $t$ is a control parameter
such as the temperature or interaction strength, and
$t_c$ is its critical value.
The stationary conditions for $\xi, \eta$, and $\zeta$ are explicitly given by
\begin{align}
&2a\xi + 3b\eta(\xi^2-\eta^2) + 4c\xi(\xi^2+\eta^2) = 0,
\\
&a\eta - 3b\xi\eta\zeta + 2c\eta(\xi^2+\eta^2) = 0,
\\
&2a'\zeta + b\xi(\xi^2-3\eta^2) = 0.
\end{align}
Since the system has a high symmetry in the orbital and sublattice spaces,
solutions can be classified into two groups for the AFO state,
which is schematically illustrated in Fig.~\ref{fig:tp}.
One is characterized by the condition $\xi=0$ and $\eta\neq 0$,
which is equivalent to $(\xi,\eta)=(0,\pm r)$ and
$(\pm\tfrac{\sqrt 3}{2}r,  \pm\tfrac{1}{2}r)$ with a positive constant $r$.
In this case, the constant $b$ is irrelevant in Eq.~(\ref{F}) and $\zeta=0$.
We then obtain the solution as
\begin{align}
  \eta &= \pm \sqrt{\frac{-a}{2c}} \propto |t_c - t|^{1/2}.
\end{align}
The \gAFO state is realized in this case.
The condensation energy $\varDelta F = F - F_0$ is given by
\begin{align}
\varDelta F_{AFO}^{(\eta)} = - \frac{a^2}{4c} <0.
\end{align}

The other group is given by the solutions
$(\xi,\eta)=(\pm r,0)$ and $(\pm\tfrac{1}{2}r,\pm\tfrac{\sqrt 3}{2}r)$.
When one fixes $\eta=0$, the solutions are given by
\begin{align}
\xi &\simeq \pm \sqrt{\frac{-a}{2c}} \propto |t_c-t|^{1/2},\\
\zeta &\simeq \mp \frac{b}{2a'}\left( \frac{-a}{2c} \right)^{3/2} \propto |t_c-t|^{3/2},
\end{align}
where we have expanded the expressions by a small parameter $a$ ($<0$)
near the continuous transition point.
Thus, the charge moment as well as the orbital moment are simultaneously induced.
Note that the critical exponents for $\xi$ and $\zeta$ are different from each other,
which implies that the CDW state is only induced by the realization of the orbital symmetry broken state (denoted by AFO($\xi$) in Fig.~\ref{fig:tp}).
In this case, the corresponding condensation energy is given by
\begin{align}
\varDelta F_{AFO-CDW}^{(\xi)} =- \frac{a^2}{4c} + \frac{b^2a^3}{32a'c^3} + O(a^4).
\end{align}
In the positive $c$ case, $\Delta F_{AFO-CDW}^{(\xi)} < \Delta F_{AFO}^{(\eta)}$.
Therefore, we can say that the CDW state is more stable than
the \gAFO state.
On the other hand, if the paramter $c$ in the fourth-order term is negative,
the genuine AFO solution may be realized.
In this case, the
sixth-order term is relevant in the free energy and
the corresponding transition should be of first order, or discontinuous.
These results are consistent with the numerical results in Fig.~\ref{PD}.
Thus the simple Landau theory explains several aspects of the phase transitions
of the orbital orders in the three-orbital Hubbard model.
We note that the analysis can be used for the system with spin-flip and pair hopping, since only the permutation symmetries of orbital and sublattice indices are invoked in the formalism.

\section{Comment on the triple point}\label{sec:comment-triple-point}
In the numerically obtained phase diagram, there are triple points
at which phase boundaries terminate.
Here we comment on the nature of the triple point based on a simple argument and investigate how it should look like in general.

\begin{figure}[t]
\begin{center}
\includegraphics[width=85mm]{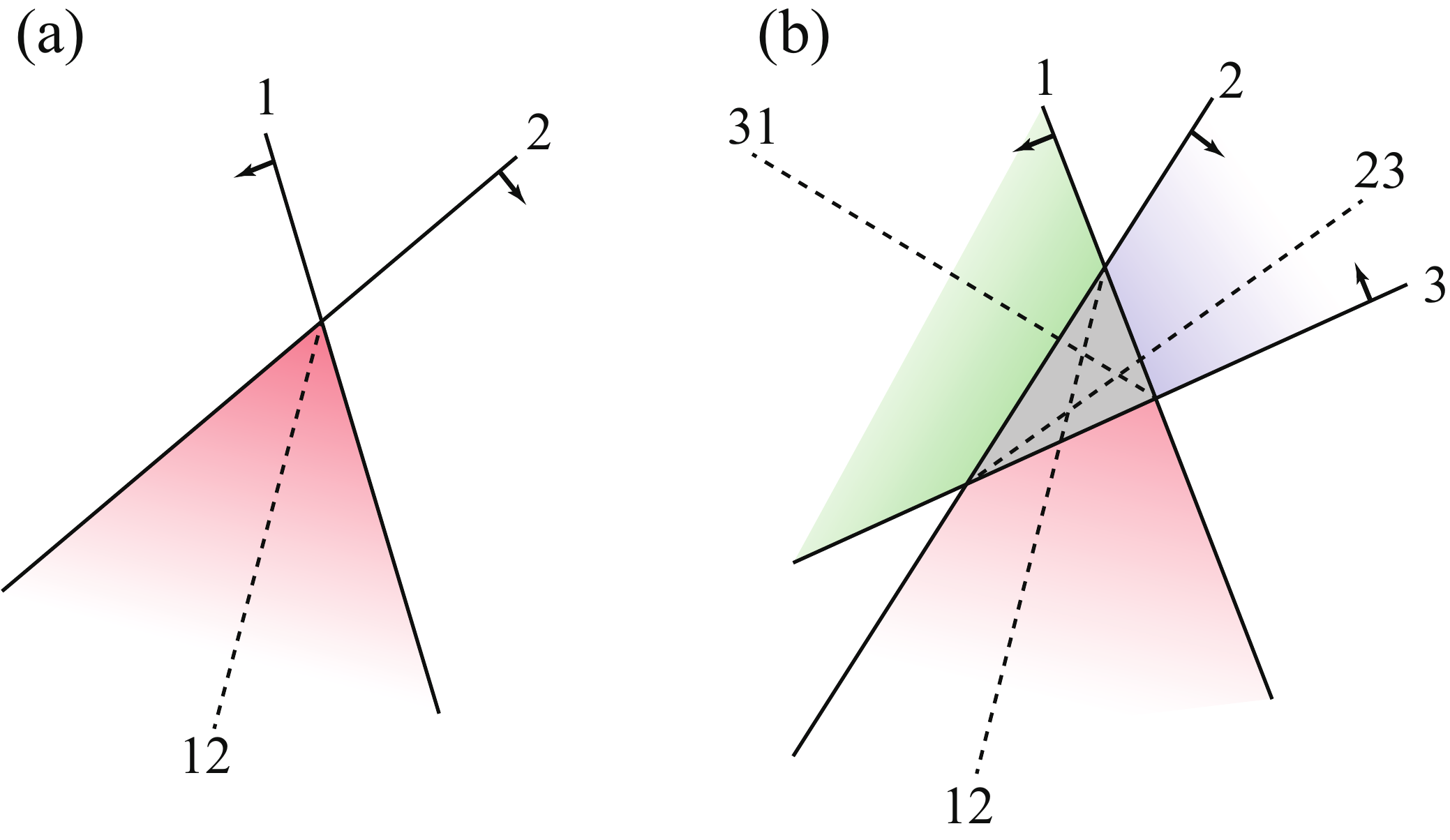}
\caption{
Schematic illustrations for (a) one and (b) three termination points of the first-order transition lines.
In the colored regions, two or more metastable solutions exist.
}
\label{fig:tp_appendix}
\end{center}
\end{figure}

First, let us consider a first-order transition between two states as shown in Fig. \ref{fig:tp_appendix}(a).
The solid lines delimit the (metastable) regions, with the arrows indicating the regions where the solution exists.
The first-order transition line, where the two free energies become identical, is shown by the dotted line.
In the colored region, the two states can exist as metastable states.

We extend this consideration to the situation with three states.
If there are three first-order transition lines, the temination points should generically look like Fig.~\ref{fig:tp_appendix}(b).
Inside the gray triangle, the three states can exist as metastable states.
Hence, in general, the first-order lines do not terminate at a single point, and instead of a triple point one should find an extended coexistence region as shown in the figure.
The phase transition lines can terminate at a single point if the three (metastable) boundaries cross at a single point.

 \bibliography{./refs}

\end{document}